\documentclass[superscriptaddress,showpacs,aps,twocolumn,prl,floatfix]{revtex4}
\usepackage{bm,amsmath,amssymb}
\usepackage{graphicx,color}

\begin{document}

\title{Theory of differential conductance of Co on Cu(111) including Co $s$ and $d$ orbitals, and surface and bulk Cu states}

\author{J. Fern\'andez}
\affiliation{Centro At\'{o}mico Bariloche, Comisi\'{o}n Nacional
de Energ\'{\i}a At\'{o}mica, 8400 Bariloche, Argentina}
\affiliation{Instituto Balseiro, Comisi\'{o}n Nacional
de Energ\'{\i}a At\'{o}mica, 8400 Bariloche, Argentina}
\affiliation{Consejo Nacional de Investigaciones Cient\'{\i}ficas y T\'ecnicas,
1025 CABA, Argentina}

\author{P. Roura-Bas}
\affiliation{Centro At\'{o}mico Bariloche, Comisi\'{o}n Nacional
de Energ\'{\i}a At\'{o}mica, 8400 Bariloche, Argentina}
\affiliation{Instituto Balseiro, Comisi\'{o}n Nacional
de Energ\'{\i}a At\'{o}mica, 8400 Bariloche, Argentina}
\affiliation{Consejo Nacional de Investigaciones Cient\'{\i}ficas y T\'ecnicas,
1025 CABA, Argentina}

\author{A. A. Aligia}
\affiliation{Centro At\'{o}mico Bariloche, Comisi\'{o}n Nacional
de Energ\'{\i}a At\'{o}mica, 8400 Bariloche, Argentina}
\affiliation{Instituto Balseiro, Comisi\'{o}n Nacional
de Energ\'{\i}a At\'{o}mica, 8400 Bariloche, Argentina}
\affiliation{Consejo Nacional de Investigaciones Cient\'{\i}ficas y T\'ecnicas,
1025 CABA, Argentina}
\email{aligia@cab.cnea.gov.ar}

\begin{abstract}
We revisit the theory of the Kondo effect observed by a
scanning-tunneling microscope (STM) for 
transition-metal atoms (TMAs) on noble-metal surfaces, including $d$ and $s$ orbitals of the 
TMA, surface and bulk conduction states of the metal, and their hopping
to the tip of the STM.
Fitting the experimentally observed STM differential conductance for Co on Cu(111) including both, 
the Kondo feature near the Fermi energy and the resonance below the surface band, we conclude that the STM senses 
mainly the Co $s$ orbital and that the Kondo antiresonance is due to interference between states
with electrons in the $s$ orbital and a localized $d$ orbital mediated by the conduction states.
\end{abstract}



\maketitle

{\em Introduction}.
The detailed understanding of the interactions of a localized spin on a metallic surface
with extended states are essential in promising quantum technologies, such as spintronics \cite{wolf} where miniaturization reaches the atomic level. Several systems in which transition-metal atoms (TMAs), such as Co, Ti or Cr, or molecules containing TMAs were deposited on noble-metal surfaces have been studied with scanning-tunneling microscopy 
\cite{coau,mano,jam,naga,knorr,wahl04,limot,wahl,neel,vita,wahl2,franke,choiPRL,mina,zhang,
karan,choi,esat,iancu,giro,hira,li,coag,maria,iwata}.
The TMAs have a localized spin in the $d$ shell, in which there are strong correlations,
which are included in most theoretical treatments 
\cite{uj,meri,mirages,lin,trimer,baru,frank,morr}.

An ubiquitous phenomenon present in these systems is the Kondo effect. This effect is one of the most paradigmatic phenomena in condensed matter systems \cite{hewson-book}. In its simplest form, it is characterized by the emergence of a many-body singlet at temperatures below the characteristic Kondo temperature $T_K$, formed by the localized spin and the spin of the conduction electrons 
near the Fermi level. As a consequence, the 
spectral density of the $d$ electrons shows a resonance at the Fermi energy. This resonance has the effect of pushing the conduction states away from the Fermi level and their spectral density shows a dip 
or Kondo antiresonance \cite{uj}. This effect is easily obtained using
equations of motion for the Green functions of 
the conduction electrons \cite{mirages}.

The observed differential conductance $dI/dV$ has been usually interpreted
using a phenomenological expression derived by Fano \cite{fano} for a non-interacting system, 
which takes into account the interference between localized and conduction states. 
According to the interpretation nowadays, the 
shape of $dI/dV$ near zero voltage is determined by the ratio of 
the hoppings of the STM tip 
to the $d$ and to the conduction electrons \cite{mirages,morr}. If the former dominates, the differential conductance represents the spectral density of the $d$ electrons and a peak is observed. Instead if 
the hopping of the STM tip 
to the conduction states dominates, a Fano-Kondo antiresonance is observed as a consequence of the corresponding dip in the conduction spectral density of states \cite{mirages,morr,uj,note}. 

In contrast to other noble metal surfaces, the (111) surfaces host a surface conduction band of Schockley states at the Fermi energy $\epsilon_F$ \cite{limot,hul,yan}. The corresponding density of states is constant and begins nearly 70 (450) meV 
below $\epsilon_F$ for Ag (Cu or Au).
Recent experiments by two different groups show the relevance of surface states in the Fano-Kondo antiresonance observed for Co on Ag(111) \cite{li,coag}. 

A crucial experiment that motivates our study is the observation in the differential conductance 
of a resonance below the bottom of the conduction band, present when either magnetic or non-magnetic TMAs are added on noble-metal surfaces \cite{coau,limot,ols}. A simple theoretical model indicates that $dI/dV$ corresponds to the spectral density of a single atomic level, most likely an $s$ one of the TMA, that hybridizes with surface and bulk states \cite{limot}. Due to the spatial extension of valence $s$ states of TMAs, it is very reasonable to expect that they have a large hopping to the conduction states of the metal and also to the STM tip. 
In fact, the fit assumes implicitly that the hopping between the STM tip and the $s$ orbital is larger than the corresponding ones between the STM tip and the conduction electrons.
However, most of the previous studies of the Kondo line shape neglect the $s$ states. 
Furthermore, while models exist that fit 
the observed $dI/dV$ near the Fermi level (Kondo effect) and near the bottom of the surface conduction 
band {\em separately}, a unified theory for both features is lacking so far. Our work closes this gap. 

In this Letter, we provide a theory for the differential conductance $dI/dV$ for Co on Cu(111) 
from voltage values below the onset of the surface band to positive values, including those 
corresponding to the resonance below the bottom of the conduction band and the Kondo antiresonance \cite{note2}. Fitting both features together puts severe constraints on the hybridization between the $d$ state and the extended conduction states and on the hopping between the tip and the different states. We find that the tip senses mostly the $s$ state, 
which gathers information on the resonance below the onset of 
the surface band and the Kondo antiresonance through its hybridization with the extended conduction states. 

{\em Model and formalism}.

\begin{figure}[tbp]
\begin{center}
\includegraphics[clip,width=\columnwidth]{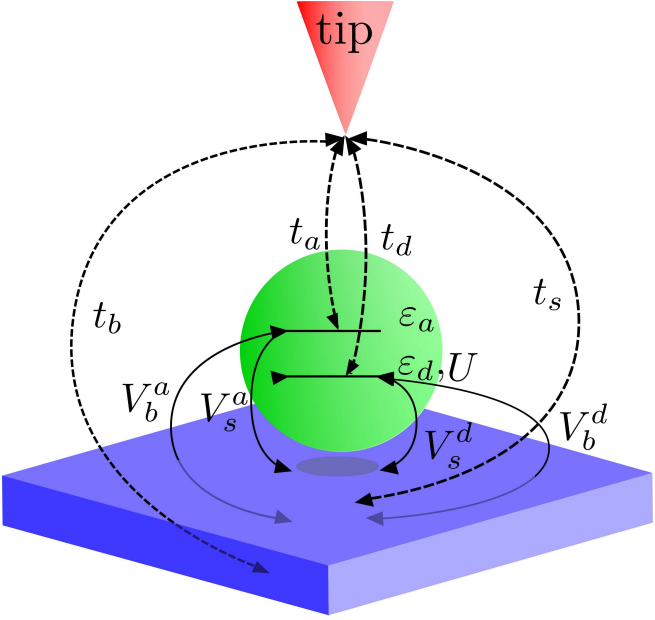}
\end{center}
\caption{Sketch of the system. The Co atom is described by a  non-interacting level $a$ representing an $s$ orbital and a $d$ level intra-orbital Coulomb  repulsion $U$. Both 
levels hop to the bulk and surface conduction states.}
\label{fig1}
\end{figure}

A sketch of the system is represented in Fig. \ref{fig1}. The Hamiltonian is

\begin{eqnarray}
 H &=& \sum_\sigma \varepsilon_a a^\dagger_\sigma a_\sigma 
 + \sum_\sigma \varepsilon_d d^\dagger_\sigma d_\sigma 
 + U d^\dagger_\uparrow d_\uparrow d^\dagger_\downarrow d_\downarrow \notag \\
 &+& \sum_{c=s,b} \sum_{k \sigma}\varepsilon_{ck} c^\dagger_{k \sigma} c_{k \sigma}
+ \sum_{c=s,b} \sum_{k \sigma} \left( V_{c}^a a^\dagger_\sigma c_{k\sigma} + \textmd{H.c.} \right), \notag \\
 &+& \sum_{c=s,b} \sum_{k \sigma} \left( V_{c}^d d^\dagger_\sigma c_{k\sigma} + \textmd{H.c.} \right). \label{ham1}
\end{eqnarray}

The first three terms represent an $s$ (denoted by $a$) and a $d$ orbital of the Co atom, 
and the interaction between two $d$ electrons. The forth term describes the two conduction bands
corresponding to bulk ($b$) and surface ($s$) extended states. The remaining terms describe the hybridization between Co and conduction states. An analysis based on symmetry indicates 
that the $d$ orbital corresponds to the $3z^2-r^2$  one \cite{note2} and that the $s$ 
orbital has an admixture with the $p_z$ one \cite{frank,note2} which lies at higher energy. 

A model containing $s$ and $d$ orbitals has been studied in Ref. \onlinecite{frank}, but the 
surface states were not included, and therefore the resonance in $dI/dV$ for $eV$ 
near the bottom of the surface
band cannot be described. In addition, the authors obtained a peak instead of a dip for the feature near $V=0$.

In the tunneling regime (as opposed to the contact regime \cite{lorente,asym}) of the STM, 
the differential conductance $dI/dV$ is proportional to the spectral density of a state $h_\sigma^\dagger$ which consists of a linear combination of all 
local and extended states with a coefficient proportional to the 
corresponding hopping to the tip \cite{mirages}:

\begin{equation}
\frac{dI(V)}{dV} \propto \rho_{h \sigma} (eV) = -\frac{1}{\pi}\textmd{Im}\langle \langle h_\sigma; 
h_\sigma^\dagger \rangle \rangle_{\omega=eV} ,
\label{didv}
\end{equation}
Assuming a local hopping of the tip with the different states, 
the state $h_\sigma^\dagger$ for spin $\sigma$
can be written as

\begin{equation}
h_\sigma^\dagger = t_{a} a_\sigma^\dagger + t_{d} d_\sigma^\dagger + 
t_{s} s_{\sigma}^\dagger(r_t) + t_{b} b_{\sigma}^\dagger(r_t),
\label{h}
\end{equation}
where $c_{\sigma}(r_t)$ denotes the operator of (surface $c=s$ or bulk $c=b$) conduction states  
at the Wannier function below the tip
and $t_{\mu}$ ($\mu=a$, $d$, $s$ or $b$) are proportional to the hopping between the tip and the different states.

Alternatively Eq. (\ref{h}) can be derived from the formalism of Meir and Wingreen \cite{meir} assuming that the presence of the STM tip does not disturb the rest of the system and 
that the whole potential difference falls between the tip and the rest of the system \cite{capac}.

We assume that the tip is located just above the impurity, see Fig. \ref{fig1}, ($r_t=r_{\textmd{imp}}$, denoting $r_t$ and $r_{\textmd{imp}}$ the position of the tip and the adatom on the surface respectively). A generalization 
to $r_t \neq r_{\textmd{imp}}$ is straightforward \cite{mirages}.
The electrons of the tip can hop to both TMA levels and the conduction states as sketched in Fig. \ref{fig1}. 
Using the equations of motion, we can write the Green function of the mixed state as

\begin{equation}
\langle \langle h_\sigma; h_\sigma^\dagger \rangle \rangle_\omega = \sum_c t_{c}^2 G_c^0(\omega) + F(\omega)
\label{gh}
\end{equation}
with
\begin{equation}
F(\omega) = \sum_\xi \tilde{t}_{ \xi}^2 \langle \langle \xi_\sigma; \xi_\sigma^\dagger \rangle \rangle_\omega + 
2 \tilde{t}_{ d}\tilde{t}_{ a} \langle \langle d_\sigma; a_\sigma^\dagger \rangle \rangle_\omega
\label{f}
\end{equation}
where $\xi=a$ or $d$ denotes the TMA orbitals and $\tilde{t}_{ \xi}$ is defined as
\begin{equation}
\tilde{t}_{ \xi} = t_{ \xi} + \sum_c t_{c} G_c^0(\omega) V_c^\xi
\label{vtilde}
\end{equation}

{\em Outline of the calculations}.
Our first step was to map the Hamiltonian Eq. (\ref{ham1}) into 
a simpler Anderson model which hybridizes the localized $d$ state
with a single band of non-interacting states which includes the
surface and bulk conduction states as well as the $s$ state. 
The details are contained in Section 2 of the 
supplemental material (SM) \cite{sm}.
This non-interacting band has energy dependent 
density of states and hybridization with the $d$ state.

This model is solved using the numerical renormalization group \cite{zitko14}, from which we obtain the Green function of the 
$d$ states 
$\langle \langle d_\sigma; d_\sigma^\dagger \rangle \rangle_\omega$
with high accuracy.

Finally, using equations of motion, the Green function entering 
the differential conductance 
[Eqs. (\ref{didv}), (\ref{h}), (\ref{gh}), (\ref{f}) and (\ref{vtilde})]  
can be exactly expressed in terms of 
$\langle \langle d_\sigma; d_\sigma^\dagger \rangle \rangle_\omega$, 
as explained in the SM \cite{sm}.

{\em Parameters of the Hamiltonian}.

We take constant hybridizations and unperturbed densities 
of conduction states. The surface density of states per spin 
in the absence of the Co atom,
corresponds to two-dimensional free electrons and is known to be
constant \cite{limot,hul}. We include lifetime effects in the lower band edge, following the experimental adjustment made by Limot \textit{et al.} \cite{limot}. Details are in the SM \cite{sm}.
Since the Fermi wavelength of surface electrons is much larger than
the atomic size, the corresponding hybridizations should have 
very weak $k$-dependence. The energy dependence of the 
unperturbed bulk density 
of states and the hybridizations with Co $s$ and $d$ states 
is expected to be weak in the range of energies of interest
and does not affect our main conclusions.

We choose the origin of energies at the Fermi level $\varepsilon_F=0$. We have taken  
$\varepsilon_a=0.33$ eV, $V_b^a=-1.41$ eV and $V_s^a=-1.46$ eV from Ref. \onlinecite{limot}
and $\varepsilon_d=-0.8$ eV from Ref. \onlinecite{mirages}. The results near the Kondo feature are rather insensitive to $\varepsilon_d$ if
the ratios $\Delta_c^d/\varepsilon_d$ are kept constant, 
where $\Delta_c^\xi=\pi \rho_c \left(V_c^\xi\right)^2$, $c=b$ or $s$ and
$\xi=d$ or $a$.
From the splitting between the positions of the majority
and minority peaks in the spectral density of Co states on Ag(111) obtained by first-principles calculations, $U=1.6$ eV is estimated \cite{tesis} [we expect a similar $U$ for Co on Cu(111)].
 The width of the Kondo feature is basically determined by $\Delta_b^d+\Delta_s^d$, which acts as a 
constraint on the parameters. The amplitude of the 
observed Fano antiresonance at the Fermi energy decreases with 
decreasing $R=\Delta_s^d/\Delta_b^d$, and 
too small $R$ is incompatible with the experiment, Details are in the SM \cite{sm}. Taking into account recent studies in similar systems \cite{coag} we have taken $R=0.5$. 
If both the resonance near the bottom of the surface conduction band and the Kondo dip were
measured in a single experiment one could quantitatively determine $R$.

In contrast to previous works in which only the Kondo feature was fitted, 
we find that the relative {\em sign} of the different $V_c^\xi$ plays a mayor role. Because of the symmetry of the $s$ 
orbitals that form the conduction band and the Co $s$ orbital, one expects that $V_c^a <0$. Instead the sign of $V_c^d$
is difficult to predict on general physical grounds. Our results indicate $V_b^d < 0, V_s^d >0$.

\begin{figure}[tbp]
\begin{center}
\includegraphics[clip,width=7cm]{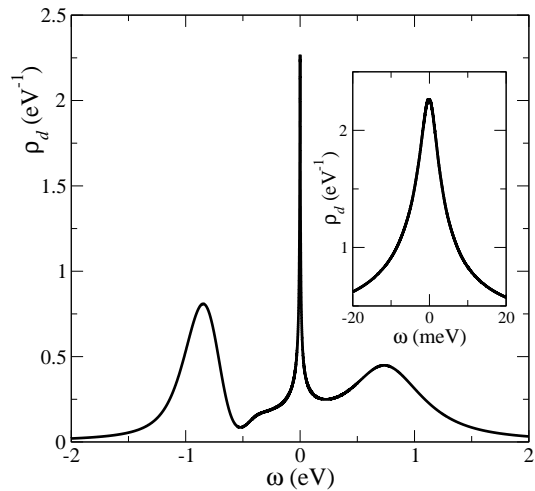}
\end{center}
\caption{Spectral density for a given spin of the $d$-orbital as a function of energy for 
$\varepsilon_d=-0.8$ eV, $U=1.6$ eV, $\varepsilon_a=0,33$ eV, 
$V_b^d=-0.50$ eV, $V_s^d=0.62$ eV, $V_b^a=-1.41$ eV and $V_s^a=-1.46$ eV at $T=4$ K,
The inset shows details of the Kondo peak near $\omega=0$.}
\label{figd}
\end{figure}

{\em Results}.
We discuss first the general features of the spectral densities of states $\rho_d(\omega)$ and $\rho_a(\omega)$ for a given spin and then
present our fits for the observed differential conductance.
The spectral density for $d$ electrons, shown in 
Fig. \ref{figd}
has the expected features 
for the impurity Anderson model, in particular a resonance at the Fermi energy, and in addition a small step
at the onset of the surface band.

\begin{figure}[tbp]
\begin{center}
\includegraphics[clip,width=\columnwidth]{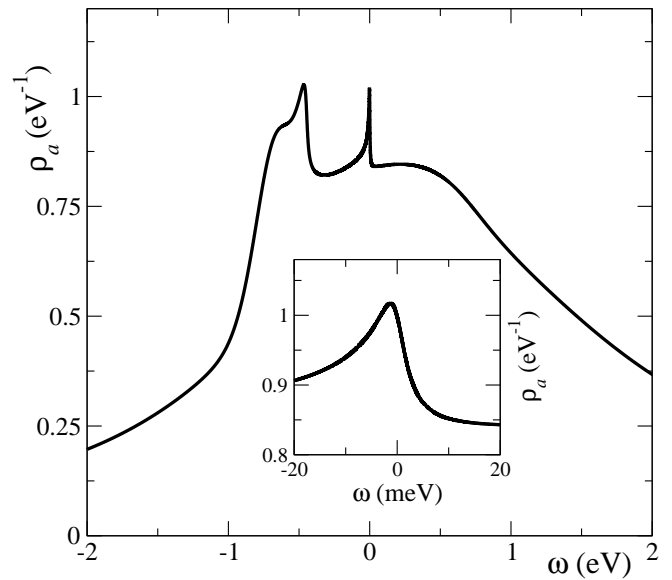}
\end{center}
\caption{Spectral density for a given spin  of the Co $s$-orbital as a function of energy for the same parameters as Fig. \ref{figd}.
The inset shows details of the peak near $\omega=0$.}
\label{figa}
\end{figure}

The spectral density for the $s$ state is displayed in Fig. \ref{figa}. The resonance
below the onset of the surface band is clearly seen. 
As a first approximation, this resonance can be understood 
as a result of the hybridization of the $s$ state with a surface bound
state, broadened by the hybridization with bulk conduction states.
This point is discussed further below.

In addition, there is also a peak at the Fermi energy. This 
is due to an effective hybridization between Co $s$ and $d$ orbitals mediated by the bulk and surface conduction bands.

\begin{figure}[tbp]
\begin{center}
\includegraphics[clip,width=\columnwidth]{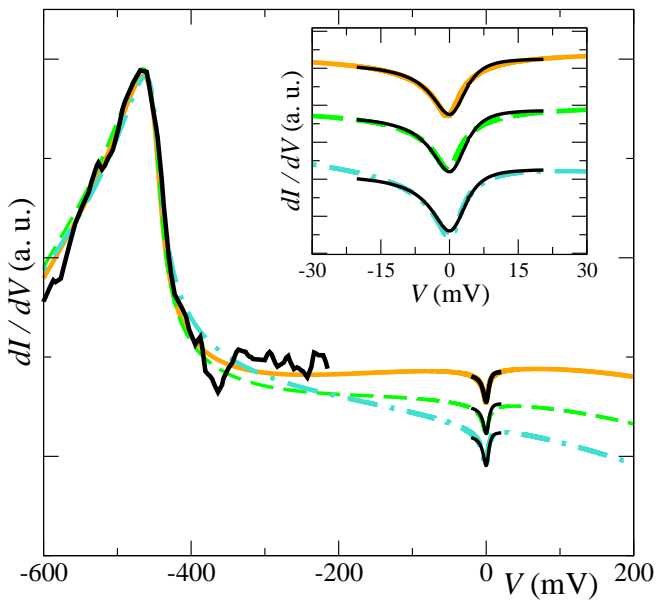}
\end{center}
\caption{Differential conductance as a function of voltage. Black solid line: experimental $dI/dV$ for Co/Cu(111). Orange solid line:  
$t_{a}=-0.978$, $t_{d}=-0.175$, $t_{s}=-0.1$ and $t_{b}=-0.04$. 
Green dashed line: 
$t_{a}=-0.959$, $t_{d}=-0.192$, $t_{s}=-0.2$ and $t_{b}=-0.08$.
Blue dashed-dot line: 
$t_{a}=-0.925$, $t_{d}=-0.185$, $t_{s}=-0.31$ and $t_{b}=-0.12$.
The inset displays the Kondo dip.}
\label{fig4}
\end{figure}

In Fig. \ref{fig4}, we compare the observed differential conductance $dI/dV$ for Co on Cu(111) \cite{limot,note} 
with our theory given by  Eq. (\ref{didv}). 
In spite of the fact that the spectral density of both $s$ and $d$ electrons has a peak 
at the Fermi energy, $dI/dV$ has a dip. This is due to the fact that the imaginary part of
the crossed Green function $\langle \langle d_\sigma; a_\sigma^\dagger \rangle \rangle_\omega $
is negative and dominates the behavior of $dI/dV$ through the last term of Eq. (\ref{f})  \cite{sm}.

Two different experiments were performed for the regions near -0.5 eV and 0 eV. Then, the corresponding experimental results
were multiplied by different factors. Beyond this uncertainty, the comparison between theory and experiment is excellent. 
The locations of the resonance below the onset of the surface band 
and the Kondo dip are well reproduced as well as the width of them. 
The parameters for the adjustment are normalized 
in such a way that $t_{a}^2+t_{a}^2+t_{s}^2+t_{b}^2=1$. For small values of $t_{a}$, the resonance near -0.5 eV cannot be fitted. An analysis of the variation of the fit with different parameters is in Section 3 of the SM \cite{sm}. The fit is practically unchanged along a line in a three-dimensional space of 
the independent $t_i$ as long as $0.925 \leq |t_{a}| \leq 0.978$. For smaller values of $|t_{a}|$, the fit deteriorates rapidly near -300 and -15 V and the magnitude of the slope between these voltages
increases. 

We note that the presence of the $s$ state is essential to reproduce
the experimentally observed $dI/dV$. While it is known that any attractive
scattering potential leads to a bound state below a two-dimensional
band, the shape of the resulting bound state in the spectral density 
of the surface states 
for a local scattering without including the $s$ state
(Fig. 8 of Ref. \onlinecite{reso}) 
is different from that observed. In Fig. S7 
of the SM \cite{sm} we compare the contribution of the $s$ state
and the surface conduction band to $dI/dV$. They are very different 
and only the former agrees with experiment.

\begin{figure}[tbp]
\begin{center}
\includegraphics[clip,width=\columnwidth]{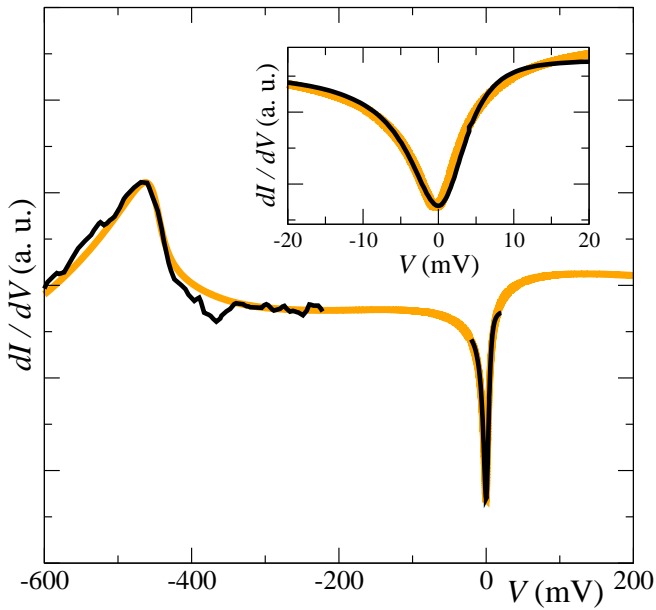}
\end{center}
\caption{Same as Fig. \ref{fig4} for $V_s^d=-1.36$ eV, $t_{a}=-0.67$, $t_{d}=-0.24$, $t_{s}=-0.67$ and $t_{b}=0.20$. 
Inset shows the Kondo dip at low energies.}
\label{fig5}
\end{figure}

In Fig. \ref{fig5} we show the best fit for a negative value of $V_s^d$. For all negative $V_s^d$, the magnitude of the 
Kondo dip is significantly larger than that of the feature below the surface band, which seems very unlikely 
in comparison with the experiments for Co on noble metal surfaces \cite{coau,limot}.

{\em Summary and discussion}.
The fact that the dominant hopping between the STM tip and the 
TMA and conduction states corresponds to the $s$ state 
($|t_{a}| \gg |t_{d}|,|t_{s}|,|t_{b}|$)
is one of the main results of this paper. 
Although this is expected from the spatial extension of 
the Co $4s$ orbital and its position near to the tip (see Fig. \ref{fig1}), this fact has been overlooked so far 
in the description of the Kondo antiresonance.
A single measurement of $dI/dV$ in the whole voltage range combined with our theory might quantify the relative importance of the surface states 
in the Kondo effect.

Other experimental observations are also consistent with our theory. $t_{a}$ is expected to be dominated by 
hopping between different $s$ orbitals, which has a $1/r$ distance dependence \cite{harri}.
Hence, when $t_{a}$ dominates, following  Eqs. (\ref{gh}), (\ref{f}), and (\ref{vtilde}), 
a $1/|r_t - r_{\textmd{imp}}|^2$ distance dependence of the differential conductance is expected,
when the tip is separated from the Co atom, as observed by Knorr {\it et al.} \cite{knorr}. 
These authors have
ascribed this dependence to a minor role of the surface states in the formation of the Kondo resonance, but
this interpretation contradicts recent experiments \cite{li,coag}.
If in our results we turn off the surface states ($V_s^d=t_{s}=0$),
we obtain a peak instead of a dip at the Fermi level, in agreement with previous
theoretical works \cite{baru,frank}.
This is what is observed for Co on Cu(100) \cite{wahl04,neel,choiPRL,vita,wahl2}, 
a surface that has not Schockley surface states.

We expect that our results will be relevant for the interpretation of other STM experiments involving transition metals 
adatoms and molecules containing magnetic transition metal atoms on metallic surfaces.

{\em Acknowledgments}.
We thank Prof. R. Berndt for helpful discussions.
We are supported by PIP 112-201501-00506 of CONICET and PICT 2013-1045, PICT-2017-2726
of the ANPCyT.


\begin{thebibliography}{99}
\bibitem{note} $dI/dV$ shows a peak near the Fermi energy for Co on Cu(100) or substituting a Cu atom of the surface \cite{wahl2,iwata}, and also for several magnetic molecules on Au(111) \cite{iancu}. Instead a dip at the 
Fermi energy is the general situation for TMAs on (111) noble metal surfaces \cite{coau,mano,jam,knorr,limot,li,coag}.
\bibitem{note2} 
The symmetry of the system corresponds to the point group $C_{3v}$ \cite{coag}.
We have chosen Co on Cu(111) for the present study because the Kondo antiresonance has a symmetric line shape
at the Fermi energy suggesting that ony one $d$ orbital
(the $3z^2-r^2$ which transforms as the trivial representation $A_1$)
participates in the Kondo effect. 
For Co on Au(111) \cite{coau} 
and Co on Ag(111) \cite{li,coag} the position of the Kondo dip is shifted to the right of the Fermi energy by an amount
of the order of half the width of the dip pointing to a double degenerate $d$ level and an SU(4) Anderson model
for the description of the system \cite{coag}. 
The $s$ and $p_z$ orbitals of Co also transform as $A_1$.
\end{thebibliography}
\end{document}